\documentclass[reprint,superscriptaddress,amsmath,amssymb,aps,prl,floatfix]{revtex4-1}
\usepackage{graphicx}
\usepackage{bm}
\usepackage{hyperref}
\usepackage{todonotes}
\usepackage{nicefrac}
\usepackage[normalem]{ulem}

\usepackage{booktabs}
\usepackage{subfigure}
\usepackage{rotating}

\definecolor{fj_color}{cmyk}{1, 0.3, 0, 0}
\definecolor{cwh_color}{cmyk}{0, 0.8, 0.8, 0}
\definecolor{hr_color}{rgb}{0.0, 0.53, 0.74}

\DeclareUnicodeCharacter{2212}{-}

\begin{document}

\newcommand{\SRO}{Sr\textsubscript{2}RuO\textsubscript{4}}

\title{The Superconductivity of \SRO{} Under \textit{c}-Axis Uniaxial Stress}

\author{Fabian Jerzembeck}
\affiliation{Max Planck Institute for Chemical Physics of Solids, N\"{o}thnitzer Str 40, 01187 Dresden, Germany}
\author{Henrik S. R{\o}ising}
\affiliation{Nordita, KTH Royal Institute of Technology and Stockholm University, Hannes Alfv\'{e}ns v\"{a}g 12, SE-106 91 Stockholm, Sweden}
\author{Alexander Steppke}
\affiliation{Max Planck Institute for Chemical Physics of Solids, N\"{o}thnitzer Str 40, 01187 Dresden, Germany}
\author{Helge Rosner}
\affiliation{Max Planck Institute for Chemical Physics of Solids, N\"{o}thnitzer Str 40, 01187 Dresden, Germany}
\author{Dmitry A. Sokolov}
\affiliation{Max Planck Institute for Chemical Physics of Solids, N\"{o}thnitzer Str 40, 01187 Dresden, Germany}
\author{Naoki Kikugawa}
\affiliation{National Institute for Materials Science, Tsukuba 305-0003, Japan}
\author{Thomas Scaffidi}
\affiliation{Department of Physics, University of Toronto, Toronto, Ontario, M5S 1A7, Canada}
\author{Steven H. Simon}
\affiliation{Rudolf Peierls Center for Theoretical Physics, Oxford OX1 3PU, United Kingdom}
\author{Andrew P. Mackenzie}
\affiliation{Max Planck Institute for Chemical Physics of Solids, N\"{o}thnitzer Str 40, 01187 Dresden, Germany}
\affiliation{Scottish Universities Physics Alliance, School of Physics and Astronomy, University of St. Andrews, St. Andrews KY16 9SS, United Kingdom}
\author{Clifford W. Hicks}
\affiliation{School of Physics and Astronomy, University of Birmingham, Birmingham B15 2TT, United Kingdom}
\affiliation{Max Planck Institute for Chemical Physics of Solids, N\"{o}thnitzer Str 40, 01187 Dresden, Germany}

\date{\today}

\begin{abstract}

Applying in-plane uniaxial pressure to strongly correlated low-dimensional systems has been shown to tune the electronic structure dramatically. 
For example, the unconventional superconductor \SRO{} can be tuned through a single Van Hove singularity which results in a strong enhancement of both $T_\text{c}$ and $H_\text{c2}$.
Out-of-plane ($c$ axis) uniaxial pressure is expected to tune the quasi-two-dimensional structure even more strongly, causing it to approach two Van Hove singularities simultaneously. 
Here we achieve a record value of $3.2$~GPa compression along the $c$ axis of \SRO{}.
Although the rise in $H_\text{c2}$ shows that we are indeed approaching the van Hove points, $T_\text{c}$ is suppressed, a result that contradicts expectations based on simple two-dimensional models.  
As a first attempt to take the third dimension into account, we present three-dimensional calculations in the weak interaction limit, and discuss the extent to which they are consistent with observation.
Our experimental results highlight the importance of out-of-plane effects in low-dimensional systems in general and provide new constraints on theories of the pairing interaction in \SRO{}.
\end{abstract}

\maketitle

\section{Introduction}

\SRO{} is a famous exemplar of unconventional superconductivity, due to the quality of the available samples and the precision of knowledge about its normal state, and because the origin of its superconductivity remains unexplained in spite of strenuous effort~\cite{Maeno94_Nature, Mackenzie03_RMP, Maeno12_JPSJ, Mackenzie17_npj}.  
No proposed order parameter is able straightforwardly to account for all the existing experimental observations. 
The greatest conundrum is posed by evidence that the order parameter combines even parity~\cite{Pustogow19_Nature, Ishida20_JPSJ, Chronister21_PNAS, Petsch20_PRL} with time reversal symmetry breaking~\cite{Luke98_Nature, Grinenko21_NatPhys, Xia06_PRL}. 
This combination of properties implies, if there is no fine tuning, that the superconducting order parameter is $d_{xz} \pm id_{yz}$~\cite{Grinenko21_NatComm}. 
Under conventional understanding, this is not expected because the horizontal line node at $k_z=0$ implies interlayer pairing, while the electronic structure of \SRO{} is highly two-dimensional~\cite{Bergemann03_AIP, Ohmichi00_PRB}.  

This puzzle has led to substantial theoretical activity. Two recent proposals are $s \pm id$~\cite{Romer19_PRL, Romer21_arXiv} and $d \pm ig$~\cite{Kivelson20_npj, Wagner20_PRB} order parameters, which require tuning to obtain $T_\text{TRSB} \approx T_\text{c}$ (where $T_\text{TRSB}$ is the time reversal symmetry breaking temperature), but avoid horizontal line nodes. 
A mixed-parity state~\cite{Scaffidi20_arxiv} and superconductivity that breaks time reversal symmetry only in the vicinity of extended defects~\cite{Willa21_PRB} have been proposed to account for the absence of a resolvable heat capacity anomaly at $T_\text{TRSB}$~\cite{Li21_PNAS}.
Interorbital pairing through Hund's coupling is also under discussion~\cite{Suh20_PRR, Clepkens21_PRR, Gingras19_PRL, Ramires19_PRB}; this mechanism could yield $d_{xz} \pm id_{yz}$ order without interlayer pairing.  
Thermal conductivity and quasiparticle interference data, on the other hand, have been interpreted as evidence for a single-component, $d_{xy}$ or $d_{x^2-y^2}$ gap~\cite{Sharma20_PNAS, Hassinger17_PRX}.

Uniaxial stress has become an important probe of the superconductivity of \SRO{}. When stress is applied along the $[100]$ direction, the largest Fermi surface sheet (the $\gamma$ sheet--- see Fig.~\ref{fig:DFT}) distorts anisotropically, and undergoes a Lifshitz transition from an electron-like to an open geometry at $-0.75$~GPa (where negative values denote compression)~\cite{Barber19_PRB}. 
$T_\text{c}$ increases from 1.5~K in unstressed \SRO{} to 3.5~K, while the $c$-axis upper critical field $H_\text{c2}$ increases by a factor of twenty~\cite{Steppke17_Science}. 
It is difficult to obtain such a strong $H_\text{c2}$ enhancement without a gap that is large at the point in $k$-space where the transition occurs and the Fermi velocity falls nearly to zero --- this would be the $Y$ point for compression along $[100]$, for example.
In a two-dimensional picture, this point is parity-invariant, so the gap of odd-parity order parameters must vanish there, and the critical enhancement was therefore an early indication that the order parameter had to be even parity \cite{Steppke17_Science}. 

Naively, then, $T_\text{c}$ and $H_\text{c2}$ might be expected to rise even further under compression along the $c$ axis. 
This raises the energy of the $d_{xz}$ and $d_{yz}$ bands relative to the $d_{xy}$ band, and the resulting transfer of carriers expands the $\gamma$ sheet, pushing it towards a Lifshitz transition from an electron-like to a hole-like geometry~\cite{Burganov16_PRL}. 
This transition occurs at both the $X$ and $Y$ points --- see Fig.~\ref{fig:DFT}(c-d) --- so the increase in the Fermi-level density of states (DOS) as it is approached is expected to be larger than for the electron-to-open Lifshitz transition induced by in-plane stress. 
This electron-to-hole transition has been approached, and crossed, in thin films through epitaxial strain, and in bulk crystals by substitution of La for Sr~\cite{Kikugawa04_PRB_1, Kikugawa04_PRB_2, Shen07_PRL}, but in both cases the superconductivity was suppressed by disorder.

In this paper, we report uniaxial stress experiments up to $3.2$~GPa along the $c$ axis of \SRO{}. 
This is a record value for bulk \SRO{} and was achieved by using a focussed ion beam as a novel sample preparation technique.
Measurements of $H_\text{c2}$ confirm the qualitative expectation that the $\gamma$ Fermi surface sheet is driven towards both van Hove points. 
However, $T_\text{c}$ decreases instead of increasing: Approaching the Lifshitz transition at either the $X$ or $Y$ point dramatically enhances $T_\text{c}$, while approaching both suppresses $T_\text{c}$.
This is a major surprise, and is completely inconsistent with the simple two-dimensional models for the superconductivity that have dominated the field until now.  
In a first attempt to address this issue, and include effects related to the third dimension, we present calculations in the limit of weak coupling. 
Although these show that $c$-axis compression reduces the transition temperatures of certain order parameters, no order parameter could be identified for which the effects of out-of-plane and in-plane pressure were both captured. 
This highlights the need for other classes of theory to take $c$-axis pressure explicitly into account in the ongoing quest to identify the form of the pairing interaction and order parameter of \SRO{}.

\section{Results}

\subsection{Electronic structure calculations}

We start with density functional theory calculations of \SRO{} under $c$-axis compression as a guide to the likely effects of strain on the electronic structure.
Unstrained lattice parameters were taken from the $T = 15$~K data of Ref.~\cite{Chmaissem98_PRB}. Longitudinal strain $\varepsilon_{zz}$ is taken as the independent variable, and $\varepsilon_{xx}$ and $\varepsilon_{yy}$ are set following the low-temperature Poisson's ratio from Ref.~\cite{Ghosh21_NatPhys}, which is $0.223$ for stress along the $c$ axis.
Calculations were performed as described in Ref.~\cite{Steppke17_Science}. 
We note in particular that spin-orbit coupling was treated nonperturbatively by solving the four-component Kohn-Sham-Dirac equation~\cite{Eschrig04_book}, the calculation was done in the local density approximation, and, due to proximity of a Van Hove singularity to the Fermi level, calculations were carried out on a fine-scale mesh in $k$ space.  The apical oxygen position was relaxed. 

\begin{figure}[ptb]
\includegraphics{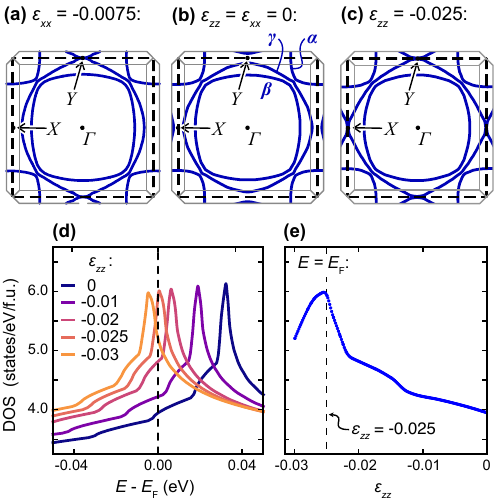}
\caption{\textbf{(a--c)} $k_z = 0$ slices through the calculated Fermi surfaces of \SRO{} at the indicated strains. The heavy dashed lines indicate the zone of the RuO$_2$ sheet, and the thin gray lines the 3D zone of \SRO{}. $X$ and $Y$ label high-symmetry points of the RuO$_2$ zone. To simplify discussion, we refer throughout this paper to the $X$ and $Y$ points defined in the 2D zone. \textbf{(a)} The electron-to-open Lifshitz transition induced by in-plane strain, from Ref.~\cite{Steppke17_Science}. Note that, although it was calculated to occur at $\epsilon_{xx} = −0.0075$, it was observed experimentally to occur at $\epsilon_{xx} \approx −0.0044$~\cite{Barber19_PRB}. \textbf{(b)} Unstressed \SRO{}. \textbf{(c)} The electron-to-hole Lifshitz transition predicted to occur under $c$-axis compression. \textbf{(d)} Calculation of the Fermi-level DOS against energy for a series of strains $\varepsilon_{zz}$. \textbf{(e)} Calculated Fermi-level DOS against $\varepsilon_{zz}$.}
\label{fig:DFT}
\end{figure}

Results are shown in Fig.~\ref{fig:DFT}. The electron-to-hole Lifshitz transition is predicted to occur at $\varepsilon_{zz} = -0.025$ (where negative values denote compression). 
Low-temperature ultrasound data give a $c$-axis Young's modulus of 219~GPa~\cite{Ghosh21_NatPhys}, so this corresponds to stress $\sigma_{zz} \approx -5.5$~GPa. 
Because meV-level energy shifts can substantially alter the distance to the Lifshitz transition, there is considerable uncertainty in these values~\cite{Barber19_PRB}. 
The transition occurs approximately at the $X$ and $Y$ points of the Brillouin zone of the RuO$_2$ sheet, indicated in panel (c). We note also that while $k_z$ warping increases on all the Fermi sheets, as expected for $c$-axis compression, the $\beta$ sheet has the strongest $k_z$ warping both at $\varepsilon_{zz} = 0$ and at the Lifshitz transition.

\subsection{Experimental results}

Four samples were measured. For good stress homogeneity, samples should be elongated along the stress axis, which is a challenge for the $c$ axis because the cleave plane of \SRO{} is the $ab$ plane. A plasma focused ion beam, in which material is milled using a beam of Xe ions, was therefore used to shape the samples.  
Sample $1$ was prepared with a uniform cross section, and a large enough stress, $\sigma_{zz} = -0.84$~GPa, was achieved to observe a clear change in $T_\text{c}$. 
To go further, the other samples were all sculpted into dumbell shapes, with the wide ends providing large surfaces for coupling force into the sample. 
FIB microsructuring has been shown to be a powerful tool for studying stress-strain curves in micropillars of CaFe$_2$As$_2$ \cite{Sypek17_NatComm}.
Here our goals were slightly different because we wanted to retain sufficient sample volume to enable high-precision magnetic susceptibility measurements.
For measurement of $T_\text{c}$ in the neck portion, two concentric coils of a few turns each were wound around the neck. 
Samples 1 and 4 also had electrical contacts, for measurement of the $c$-axis resistivity $\rho_{zz}$.  Photographs of the samples are shown in the Methods section.

\begin{figure}[ptb]
\includegraphics[width=85mm]{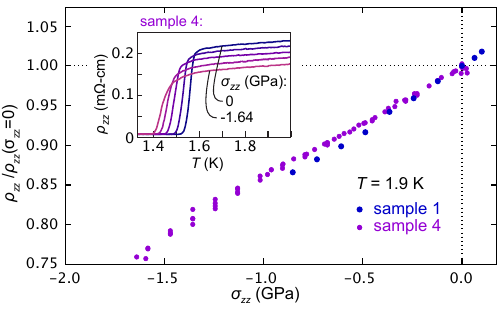}
\caption{Main panel: $c$-axis resistivity $\rho_{zz}$ versus stress $\sigma_{zz}$ at 1.9~K, normalized by its $\sigma_{zz} = 0$ value.  Note that the stress scale of sample $1$ is adjusted so that $dT_c/d\sigma_{zz}$ as measured through the Meissner effect matches that from sample 4 over the range $-0.92 < \sigma_{zz} < -0.2$~GPa. At $\sigma_{zz}=0$, $\rho_{zz}(1.9~\text{K})$ of samples 1 and 4 is $0.278$ and $0.228$~m$\Omega$-cm, respectively. Inset: $\rho_{zz}$ versus temperature for sample~4 at 0, $-0.37$, $-0.85$, $-1.25$, and $-1.64$~GPa.} \label{fig:normal_state}
\end{figure}

Sample 4 was measured in apparatus that incorporated a sensor of the force applied to the sample~\cite{Barber19_RSI}, from which the stress in the sample could be accurately determined. 
Samples 1--3 were mounted into apparatus that had a sensor only of the displacement applied to the sample, which is an imperfect measure of the sample strain because the sensor also picks up deformation in the epoxy. 
Therefore, a displacement-to-stress conversion was applied to samples 1--3 to bring the rate of change of $T_\text{c}$ over the stress range $\-0.92 < \sigma_{zz} < -0.20$~GPa into agreement with that of sample~4. 
In other words, we impose on our data an assumption that the initial rate of decrease in $T_\text{c}$ is the same in all the samples, which is reasonable because their zero-stress $T_\text{c}$'s are very similar: all are between 1.45 and 1.50~K.

\begin{figure}[ptb]
\includegraphics[width=85mm]{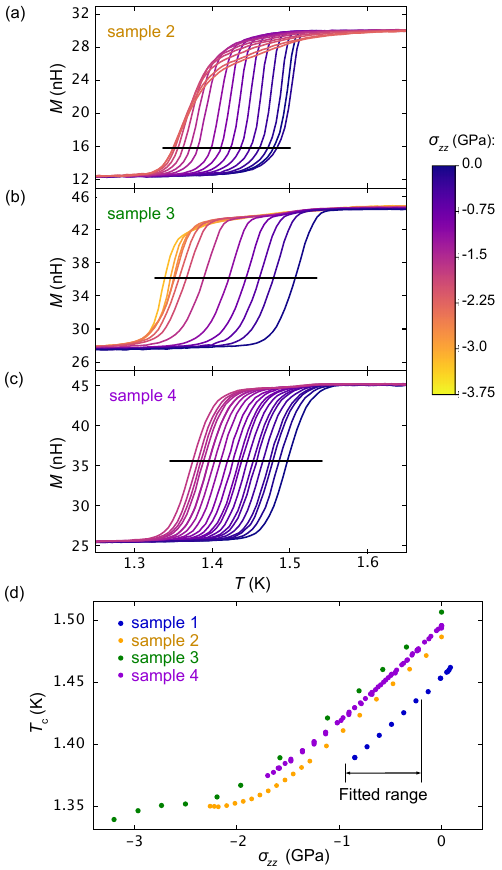}
\caption{\textbf{(a--c)} Raw data: mutual inductance $M$ of the sense coils versus temperature for samples 2--4, at various applied stresses $\sigma_{zz}$. The lines indicate the selected thresholds for determination of $T_\text{c}$. For sample $2$, under large $|\sigma_{zz}|$ a tail appears on the transitions, which we attribute to in-plane stress, and a low threshold is chosen to avoid this tail.  \textbf{(d)} $T_\text{c}$ versus stress for all the samples. The stress scales for samples 1--3 are scaled to bring $dT_\text{c}/d\sigma_{zz}$ into agreement with that of sample 4 over the stress range labeled ``fitted range,'' $-0.92$ to $-0.20$~GPa.}
\label{fig:temp_sweeps}
\end{figure}

We begin by showing resistivity data, in Fig.~\ref{fig:normal_state}. The plotted resistivities are corrected for the expected stress-induced change in sample geometry (reduced length and increased width) under an assumption that stress and strain remain proportional, and using the low-temperature elastic moduli reported in Ref.~\cite{Ghosh21_NatPhys}. 
At zero stress the resistivity of sample 4 shows a sharp transition into the superconducting state at $1.55$~K. 
This sharpness, and the fact that it only slightly exceeds the transition temperature seen in susceptibility, indicate high sample quality.  
With compression, $T_\text{c}$ decreases. The normal-state resistivity also decreases, following the general expectations that $c$-axis compression should increase $k_z$ dispersion.

We find elastoresistivities $(1/\rho_{zz})d\rho_{zz}/d\varepsilon_{zz}$, obtained with a linear fit over the range $-0.5 < \sigma_{zz} < 0$~GPa, of 37 and 32 for samples 1 and 4, respectively. Sample 4 was compressed to $-1.7$~GPa, and its resistivity does not show any major deviation from linearity over this range.
There is some scatter in the data at large compression, which may be a consequence of cracking in the electrical contacts--- we show below that the sample deformation was almost certainly elastic.

We now show the effects of interlayer compression on the superconductivity. 
In Fig.~\ref{fig:temp_sweeps} we show the dependence of $T_\text{c}$ measured through susceptibility on $c$-axis stress, for all four samples. 
Panels (a--c) show the actual transitions --- the mutual inductance $M$ of the sense coils versus temperature --- for samples 2--4. 
To check that sample deformation remained elastic, we repeatedly cycled the stress to confirm that the form of the $M(T)$ curves remained unchanged; see the Appendix for examples. 
For samples 3 and 4, the transition remained narrow as stress was applied, indicating high stress homogeneity. 
For sample 2, there was a tail on the high-temperature side of the transition, that was stronger at higher compressions. 
We attribute it to in-plane strain, possibly originating in the fact that sample 2 was not as well aligned as samples 3 and 4. A similar, though weaker, tail is also visible for sample 3.

\begin{figure}[ptb]
\includegraphics[width=85mm]{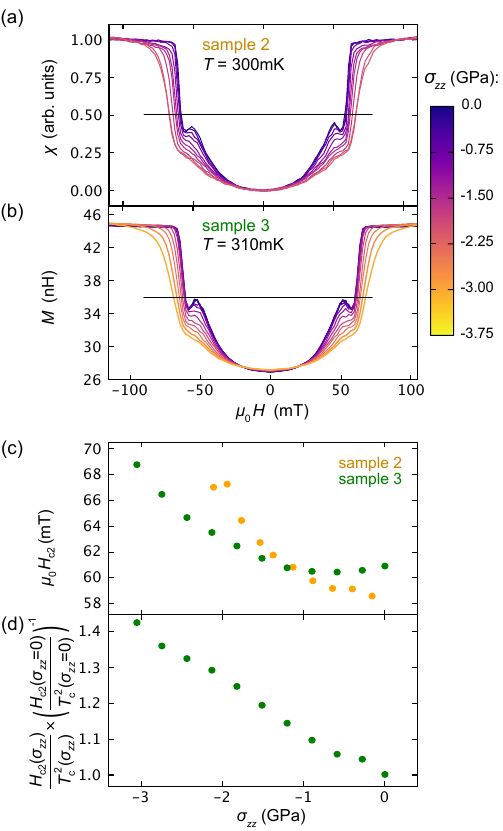}
\caption{\textbf{(a--b)} Susceptibility versus applied field $\parallel c$ for samples $2$ and $3$, at fixed temperatures of $300$~mK and $310$~mK. For
sample~$3$, the raw data are plotted. For sample~$2$, the signal magnitude shifted from run to run, so data are normalized by the readings at $0$ and $100$~mT.
This shift was probably due to motion of the coils against the sample; when they were fixed more securely to collect the data in Fig.~\ref{fig:temp_sweeps}(a),
the problem disappeared. The horizontal lines indicate the thresholds for determination of $H_\text{c2}$. \textbf{(c)} $H_\text{c2}(T \approx 0.3~\text{K})$
versus stress for samples 2 and 3.  \textbf{(d)} $H_\text{c2}/T_\text{c}^2$, normalized by its zero-stress value, versus stress for sample $3$.}
\label{fig:field_data}
\end{figure}

Panel~(d) shows $T_\text{c}$ versus stress for all the samples. $T_\text{c}$ is taken as the temperature where $M$ crosses a threshold. 
For samples 1, 3, and 4, we select a threshold at $\approx$50\% of the height of the transition, and for sample 2, 20\%, in order to minimize the influence from the high-temperature tail. 
$T_\text{c}$ is seen to decrease almost linearly out to $\sigma_{zz} \approx -1.8$~GPa. 
For sample 4 (to which, as described above, the other samples are referenced), $dT_\text{c} / d\sigma_{zz}$ in the limit $\sigma_{zz} \rightarrow 0$ is $76 \pm 5$~mK/GPa. 
The error is 6\%: we estimate a 5\% error on the calibration of the force sensor of the cell, and a 3\% error on the cross-sectional area of the sample ($155 \times 106$~$\mu$m$^2$). 

At $\sigma_{zz} \lesssim -1.8$~GPa, the stress dependence of $T_\text{c}$ flattens markedly. 
In sample 3, $T_\text{c}$ is seen to resume its decrease for $\sigma_{zz} < -3$~GPa. 
We show in the Methods section that both the flattening and this further decrease reproduce when the stress is cycled, which, in combination with the narrowness of the transitions, shows that this behavior is intrinsic, and not an artefact of any drift or non-elastic deformation in the system.

Fig.~\ref{fig:field_data} shows measurements of the $c$-axis upper critical field. $M(H)$ for samples 2 and 3 at constant temperature $T \approx 0.3$~K is shown in panels (a) and (b). 
In Fig.~\ref{fig:field_data}(c), we plot $H_\text{c2}$ versus stress, taking $H_\text{c2}$ as the fields at which $M$ crosses the thresholds indicated in panels (a--b).  
$H_{c2}$ is seen to increase as stress is applied, as generally expected when the density of states increases.  
The increase is faster for sample $2$ than sample $3$, which may be an artefact of the tail on the transition for sample 2.

For an isotropic system, $H_\text{c2} \propto (T_\text{c}/v_\text{F})^2$, where $v_\text{F}$ is the Fermi velocity, and so in panel~(d) we plot $H_\text{c2}/T_\text{c}^2$ normalized by its zero-stress value for sample~3. 
It increases by $\approx$40\% by $\sigma_{zz} = -3.0$~GPa, which, if the gap structure does not change drastically, suggests an increase in the Fermi-level DOS of $\approx$20\%.

Another feature visible in the $M(H)$ traces of Fig.~\ref{fig:field_data}(a--b) is a peak effect --- a local maximum in the susceptibility just below $H_\text{c2}$. It occurs when there is a range of temperature below $T_\text{c}$ where vortex motion is uncorrelated, allowing individual vortices to find deeper pinning sites~\cite{Yamazaki02_PhysicaC}. 
The peak is suppressed by $c$-axis compression, and it is suppressed downward rather than by being smeared horizontally along the $H$ axis, which means that it is not an artefact of a spread of $H_\text{c2}$ due to strain inhomogeneity. It could indicate stronger pinning, due to the reduction in the coherence length.

\subsection{Weak-coupling calculations}

\begin{figure}[ptb]
	\centering  
	\includegraphics[width=85mm]{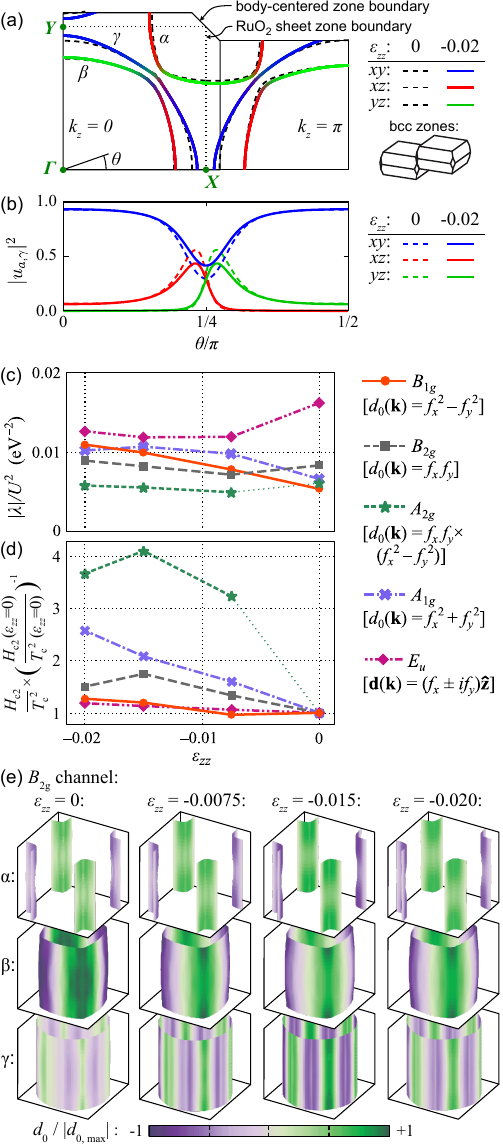}
	\caption{\textbf{(a)} Cuts through the tight-binding Fermi surfaces employed in our weak-coupling calculations, at $\varepsilon_{zz} = 0$ and
$-0.02$. The surfaces at $\varepsilon_{zz} = -0.02$ are colored by orbital content. 
	\textbf{(b)} Orbital weights on the $\gamma$ sheet at $k_z = 0$ in this model.
	\textbf{(c)} Eigenvalues as a function of $\varepsilon_{zz}$ for $J/U = 0.15$. In the legend, $f_i$ is any function that transforms as $\sin
k_i$; $d_0(\mathbf{k})$ is the gap function for the even-parity irreducible representations, and $\mathbf{d}(\mathbf{k})$ the $d$-vector for $E_u$. In the
$A_{2g}$ channel, there is a transition in the gap structure between $\varepsilon_{zz} = 0$ and $-0.0075$.
	\textbf{(d)} $H_\text{c2}(T \rightarrow 0)/T_\text{c}^2$ versus $\varepsilon_{zz}$ in each channel. 
	\textbf{(e)} Gap structure in the $B_\text{2g}$ channel.}
	\label{fig:Eigs}
\end{figure}

To further investigate the opposing trends of $T_\mathrm{c}$ and $H_\mathrm{c2}$, we supplement these results with weak-coupling calculations for repulsive Hubbard models, as developed in Refs.~\cite{KohnLuttinger65, BaranovEA92, KaganChubukov89, ChubukovEA92, BaranovChubukovEA92, HironoEA02, Hlubina99, RaghuEA10, Raghu2EA10, Scaffidi14_PRB}.
To capture possible changes in the 3D gap structure, we employ three-dimensional Fermi surfaces~\cite{Roising19_PRR}. 
These are calculated using a tight-binding model for the three bands that cross the Fermi level, that takes the form
\begin{equation}
H_0 = \sum_{\boldsymbol{k}, s} \boldsymbol{\psi}^{\dagger}_s(\boldsymbol{k}) \mathcal{H}_s(\boldsymbol{k}) \boldsymbol{\psi}_s(\boldsymbol{k}).
\label{eq:Hamiltonian} 
\end{equation}
$\boldsymbol{\psi}_s(\boldsymbol{k}) = \left[c_{xz,s}(\boldsymbol{k}), c_{yz, s}(\boldsymbol{k}), c_{xy, \bar{s} }(\boldsymbol{k}) \right]^T$, and
$\mathcal{H}_s(\boldsymbol{k})$ incorporates spin-orbit coupling, inter-orbital and intra-orbital terms as extracted from the DFT calculations.  The complete
set of tight-binding parameters retained here is given in the Methods section. In Fig.~\ref{fig:Eigs}(a), we show the tight-binding Fermi surfaces at
$\varepsilon_{zz} = 0$ and $-0.02$. In Fig.~\ref{fig:Eigs}(b), we show the orbital weight on the $\gamma$
sheet at $k_z = 0$.  Because the $\gamma$ sheet expands under $c$-axis compression, the orbital mixing around its avoided crossings with the $\beta$ sheet is
reduced, and it becomes more dominated by $xy$ orbital weight.

To $H_0$ we add on-site Coulomb terms projected onto the $t_{2g}$ orbitals~\cite{DagottoEA11} (Methods Eq.~\ref{eq:Interaction}) and study the
solutions to the linearized gap equation in the weak-coupling limit $U/t \ll 1$, where $U$ is the intraorbital Coulomb repulsion and $t$ is the leading
tight-binding term. We take the interorbital on-site Coulomb repulsion to be $U^\prime = U - 2J$, where $J$ is the Hund's coupling, and the pair-hopping Hund's
interaction $J^\prime$ to be equal to the spin-exchange Hund's interaction $J$. Under these assumptions, the remaining free parameter is $J/U$. We take $J/U =
0.15$, which is close to the value $J/U = 0.17$ found in Refs.~\cite{Mravlje11_PRL, Tamai19_PRX}. The linearized gap equation reads
\begin{equation}
\sum_{\nu} \int_{S_{\nu}} \frac{\mathrm{d} \boldsymbol{k}_{\nu}}{\lvert S_{\nu} \rvert} \bar{\Gamma}(\boldsymbol{k}_{\mu}, \boldsymbol{k}_{\nu}) \varphi(\boldsymbol{k}_{\nu}) = \lambda \varphi(\boldsymbol{k}_{\mu}),
\label{eq:GapEquation}
\end{equation}
where $\mu$ and $\nu$ are band indices, $\lvert S_{\nu} \rvert$ is the area of Fermi surface sheet $\nu$, and $\bar{\Gamma}$ is the two-particle interaction
vertex calculated consistently to order $\mathcal{O}(U^2/t^2)$. Solutions to Eq.~\eqref{eq:GapEquation} with $\lambda < 0$ signal the onset of
superconductivity, at the critical temperature $T_c \sim W \exp(-1/\lvert \lambda \rvert)$, where $W$ is the bandwidth. 

In a pseudo-spin basis each eigenvector $\varphi$ belongs to one of the ten irreducible representations of the crystal point group $D_{4h}$~\cite{SigristUeda91,
RaghuEA10}. We calculate the leading eigenvalues in four even-parity channels, $B_{1g}$, $B_{2g}$, $A_{1g}$, and $A_{2g}$ --- see the legend of
Fig.~\ref{fig:Eigs}(c--d). The $E_g$ channel --- $d_{xz} \pm id_{yz}$ --- has been found to be strongly disfavored in weak-coupling
calculations~\cite{Roising19_PRR}, and so is not considered here. We calculate only one odd-parity channel, $E_u$. The splitting among the odd-parity channels
has been found to be small in comparison with that between odd- and even-parity states for reasonable values of $J/U$ and spin-orbit coupling~\cite{WangEA20}.

The leading eigenvalues in each channel as a function of $\varepsilon_{zz}$ are shown in Fig.~\ref{fig:Eigs}(c). Although, as in Ref.~\cite{Roising19_PRR},
odd-parity order is found to be favored, calculations in the random phase approximation at similar $J/U$ tend to favor even-parity order~\cite{Romer19_PRL,
WangEA20}. A tendency towards odd-parity order appears to be a feature of calculations in the weak-coupling limit.

The weak-coupling results show a dichotomy in the strain dependence of $T_\text{c}$: $T_\text{c}$ in the channels that have symmetry-imposed nodes at the $X$
and $Y$ points ($E_u$, $A_{2g}$, and $B_{2g}$) decreases with initial $c$-axis compression. These nodes coincide with the regions of highest local density of
states, and this result is an indication that order parameters in these channels are less able to take advantage of the increase in density of states induced by
$c$-axis compression. We note, however, that under stronger compression $T_\text{c}$ increases modestly in all the channels.

We also calculate $H_\text{c2}/T_\text{c}^2$, following the procedure described in Ref.~\cite{Steppke17_Science}, and obtain the results shown in
Fig.~\ref{fig:Eigs}(d). We find that changes in $H_\text{c2}/T_\text{c}^2$ correlate closely with shifts in the gap weight onto the $\gamma$ sheet, which has
the lowest Fermi velocity. This is illustrated by the calculated gap structure in the $B_{2g}$ channel, shown in Fig.~\ref{fig:Eigs}(e); the gap structures in
the other channels are shown in Methods Fig.~\ref{fig:OPs_compare}. As stress is initially applied, gap weight shifts from the $\beta$ to the $\gamma$ sheet,
to take advantage of the increasing DOS on the $\gamma$ sheet. However, in this channel the gap changes sign across the zone boundary of the RuO$_2$ sheet, and
so at the strongest compression, where $\gamma$ sheets in adjacent zones come close to each other, the gap is suppressed on the $\gamma$ sheet.  The calculated
$H_\text{c2}/T_\text{c}^2$ follows this non-monotonic dependence: an initial increase as gap weight shifts to the $\gamma$ sheet, then a decrease as it shifts
back.  Under strong compression ($\varepsilon_{zz} < -0.015$ in this model), there is a clear dichotomy between the even-parity channels with and without nodes
along $\Gamma$-$X$ and $\Gamma$-$Y$: $H_\text{c2}/T_\text{c}^2$ decreases for the former and increases for the latter.

Although a $k_z$ dependence of the gap structure is seen in all channels, we do not find dramatic stress-induced changes in the $k_z$ dependence in any channel. 
Separately, in Fig.~\ref{fig:Eigs}(d) a very strong increase in $H_\text{c2}/T_\text{c}^2$ is found in the $A_{2g}$ channel.
This is due to a first-order change in the gap structure, indicated in the figure by a dotted line.

\section{Discussion}

The unexpected decrease of $T_\text{c}$ with $c$-axis uniaxial pressure as two Van Hove singularities are approached is the key experimental result that we report. 
It might provide a vital clue to understand the nature of the superconducting state in \SRO{}, because it is so different to the response to in-plane, $a$-axis pressure. 
To assess the significance of the differing results between the two pressure directions, we first examine the scale of the density of states increase that is being achieved in our $c$-axis pressure experiment.

An applied pressure of $\sigma_{zz} = -3.2$~GPa corresponds to a strain $\epsilon_{zz}  \approx -0.014$, about $60~\%$ of that calculated to be necessary to cause a Lifshitz transition, $-0.025$. 
However, there is considerable uncertainty in the calculated values, and so we also look at measured physical quantities to determine how far along the way to the electron-to-hole Lifshitz transition we reached. 
As noted above, $H_\text{c2}/T_\text{c}^2$, a naive estimate of the square of the density of states, increases by $40~\%$ between $\sigma_{zz} = 0$ and $-3.0$~GPa.
For $a$ axis pressure, the same $40~\%$ increase occurs at approximately $65~\%$ of the Lifshitz transition strain.
These estimates highlight the scale of the discrepancy of the behaviour of $T_\text{c}$ between the two directions of pressure: when $H_\text{c2}/T_\text{c}^2$ has risen by $40~\%$ under $a$ axis pressure, $T_\text{c}$ has increased by $0.8$~K. 
Here, with $c$-axis pressure $T_\text{c}$ drops by 0.15 K.
We are not aware of any two-dimensional model that has reproduced this behaviour. 
For example, the weak-coupling renormalization group study of Ref.~\cite{Hsu16_PRB} and functional renormalization group study of Ref.~\cite{Liu18_PRB} both predict a rapid increase in $T_\text{c}$ with approach to the electron-to-hole Lifshitz transition.

The response of $T_\text{c}$ under $c$-axis compression also allows us to resolve the stress dependence of $T_\text{c}$ into components by comparing it to the the effect of hydrostatic compression, which also suppresses $T_\text{c}$ linearly. 
We obtain the coefficients $\alpha$ and $\beta$ in the expression
\[
T_\text{c} = T_\text{c,0} + \alpha \times \frac{ \Delta V}{V} + \beta \times \left(\varepsilon_{zz} - \frac{\varepsilon_{xx} + \varepsilon_{yy}}{2}\right),
\]
where $\Delta V / V = \varepsilon_{xx} + \varepsilon_{yy} + \varepsilon_{zz}$ is the fractional volume change of the unit cell, and $\varepsilon_{zz} - (\varepsilon_{xx} + \varepsilon_{yy})/2$ is a volume-preserving tetragonal distortion. 
Refs.~\cite{Forsythe02_PRL, Grinenko21_NatComm, Svitelskiy08_PRB} report $dT_\text{c}/d\sigma_\text{hydro} = 0.22 \pm 0.02$, $0.24 \pm 0.02$, and $0.21 \pm 0.03$~K/GPa; we take $dT_\text{c}/d\sigma_\text{hydro} = 0.23 \pm 0.01$~K/GPa. 
Employing the low-temperature elastic moduli from Ref.~\cite{Ghosh21_NatPhys} to convert stress to strain, we find $\alpha = 34.8 \pm 1.6$~K and $\beta = -2.2 \pm 1.2$~K~\footnote{Under hydrostatic stress, $\sigma_{zz} = (396~\text{GPa}) \times \varepsilon_{zz}$ and $\varepsilon_{xx} = 0.814 \varepsilon_{zz}$}.
The small value of $\beta$ means that a volume-preserving reduction in the lattice parameter ratio $c/a$ would have little effect on $T_\text{c}$: the increase in density of states by approaching the electron-to-hole Lifshitz transition is balanced, somehow, by weakening of the pairing interaction. 
The challenge for theory is to understand how that weakening might take place.

In the three-dimensional weak-coupling calculations presented here, it is the $A_{2g}$ and $B_{2g}$ channels, both of which have nodes along the $\Gamma$-$X$ and $\Gamma$-$Y$ lines, that best match observations.
Due to differences between the actual and calculated electronic structures the $\epsilon_{zz} = 0$ point in the calculations should not be considered too literally as equivalent to $\epsilon_{zz} = 0$ in reality, and the key point is that it is only in the $A_{2g}$ and $B_{2g}$ channels that $T_c$ is found to decrease and $H_\text{c2}/T_\text{c}^2$ to increase over some range of strain. 
However, the $A_{2g}$ and $B_{2g}$ order parameters do not appear to be consistent with the results of in-plane uniaxial stress experiments, in which the strong increase in $H_\text{c2}$ as the electron-to-open Lifshitz transition is approached indicates that there are not nodes near the $X$ and $Y$ points. 
Although it is informative to see the extra possibilities that adding the third dimension to the calculations open up, it therefore seems that this specific approach does not solve the mystery presented by the observations.

We believe that another possibility is worthy of further theoretical attention: if the superconductivity is driven by interorbital interactions~\cite{Suh20_PRR, Clepkens21_PRR, Gingras19_PRL, Ramires19_PRB}, then $T_\text{c}$ might decrease because $c$-axis compression reduces orbital mixing. 
The superconducting energy scale is too weak to induce substantial band mixing, and so the proximity of the $\gamma$ and the $\beta$ sheets, and the resulting mixing of $xy$ and $xz$/$yz$ orbital weight over substantial sections of Fermi surface, is crucial to these models~\cite{Clepkens21_PRR}. 
We have noted that $c$-axis compression reduces this mixing, by pushing the $\gamma$ and $\beta$ sheets apart. 
In contrast, under in-plane uniaxial compression these sheets are pushed closer together along one direction and further apart along the other \cite{Sunko19_npj}. 
It would be interesting to see this investigated qualitatively in interorbital models that explicitly address the effects of both $a$- and $c$-axis compression.

In summary, we have demonstrated methods to apply uniaxial stress of multiple GPa along the interlayer axis of layered materials in samples large enough to permit high-precision magnetic susceptibility measurements.
Under such a compression, we find that $T_\text{c}$ decreases even though the Fermi-level DOS increases, in striking contrast to the effect of in-plane uniaxial stress. 
This contrast cannot be explained by current theoretical models and highlights the importance of extending current theories of superconductivity of \SRO{} to realistically incorporate the third dimension.
At a more general level, our findings motivate the use of out-of-plane stress as a powerful tool for investigation of other low dimensional strongly correlated systems.
In \SRO{}, we now see that, even though it is extremely anisotropic, working in a purely 2D approximation may not be sufficient to gain a full understanding of its superconductivity.

\begin{figure}[ptb]
\includegraphics[width=85mm]{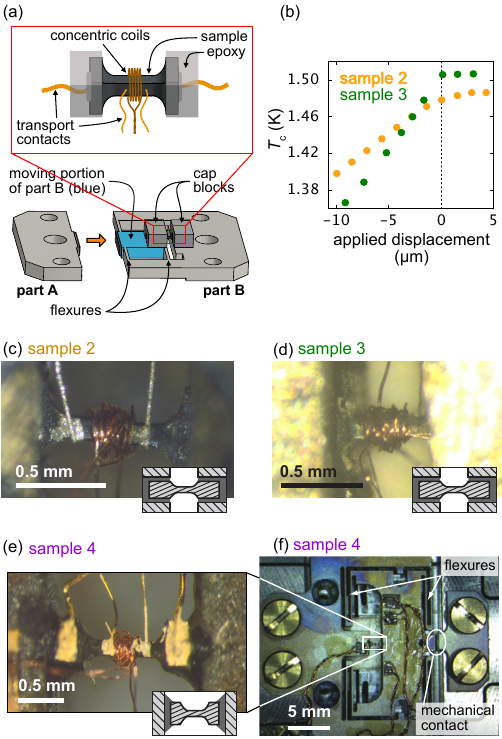}
\caption{\textbf{(a)} Schematic of the sample configuration for samples 2 and 3. These samples were sculpted into a dumbell shape with a plasma focused ion
beam. They were mounted with their ends embedded in epoxy. The cap blocks incorporate slots into which the sample fits, and the sample is compressed by bringing
parts A and B into contact. \textbf{(b)} $T_\text{c}$ versus applied displacement for samples 2 and 3. When parts A and B are brought into contact, force is
applied to the sample, and $T_\text{c}$ changes. \textbf{(c--e)} Photographs of samples 2, 3, and 4. The graphics at the lower right of each panel are schematic
cross sections: the end tabs of samples 2 and 3 were epoxied into slots, while sample 4 was sandwiched between two surfaces. \textbf{(f)} A photograph of the
sample carrier for sample 4.} \label{fig:setup}
\end{figure}

\textit{Acknowledgements.} We thank Aline Ramires and Carsten Timm for helpful discussions, Markus K\"{o}nig for training on the focused ion beam, and Felix
Flicker for help with development and running of the code. F.J., A.P.M., and C.W.H. acknowledge the financial support of the Deutsche Forschungsgemeinschaft
(DFG, German Research Foundation) - TRR 288 - 422213477 (project A10). H.S.R. and S.H.S. acknowledge the financial support of the Engineering and Physical
Sciences Research Council (UK). H.S.R. acknowledges support from the Aker Scholarship. T.S. acknowledges the support of the Natural Sciences and Engineering
Research Council of Canada (NSERC), in particular the Discovery Grant [RGPIN-2020-05842], the Accelerator Supplement [RGPAS-2020-00060], and the Discovery
Launch Supplement [DGECR-2020-00222]. This research was enabled in part by support provided by Compute Ontario (www.computeontario.ca) and Compute Canada
(www.computecanada.ca).  N.K. is supported by a KAKENHI Grants-in-Aids for Scientific Research (Grant Nos. 17H06136, 18K04715, and 21H01033), and Core-to-Core
Program (No. JPJSCCA20170002) from the Japan Society for the Promotion of Science (JSPS) and by a JST-Mirai Program (Grant No. JPMJMI18A3).  Raw data shown in
this article are available at \textit{website to be determined}.

\section{Methods}

\subsection{Experimental details}

\SRO{} samples were grown using a floating-zone method~\cite{Mao00_MRB, Bobowski19_CondMat}. 
The four samples here were taken from the same original rod, and from a portion that we verified to have high $T_\text{c}$ and a low-density of Ru inclusions; our aim in taking multiple samples was to test re-producibility in sample preparation and mounting.

\begin{figure}[ptb]
\includegraphics[width=85mm]{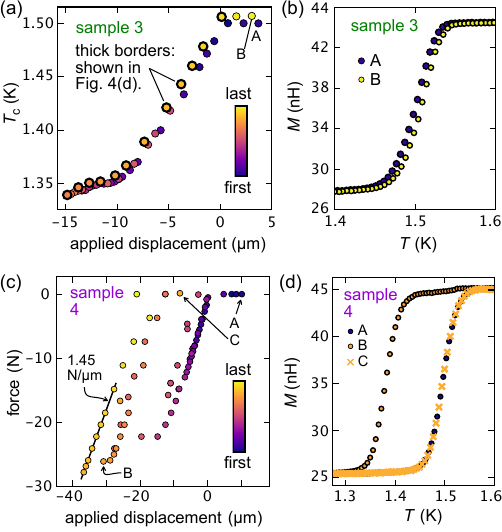}
\caption{\textbf{(a)} $T_\text{c}$ of sample 3 versus applied displacement, with the data points colored by the order they were taken.  The sequence of data points with the thick borders, during which $|\sigma_{zz}|$ was monotonically decreased, are those shown in Fig.~\ref{fig:temp_sweeps}(d). \textbf{(b)} Sense coil mutual inductance $M(T)$ at points A and B in the left-hand panel. \textbf{(c)} Force versus displacement of sample 4. \textbf{(d)} $M(T)$ at points A, B, and C in the left-hand panel.} 
\label{fig:additionalData}
\end{figure}

\begin{figure}[ptb]
\includegraphics[width=85mm]{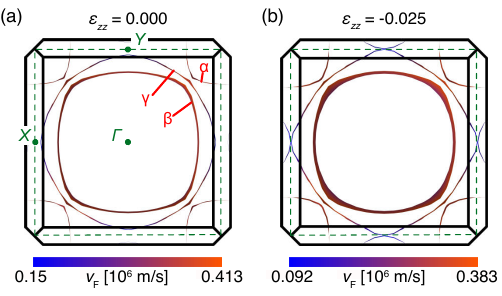}
\caption{Fermi surfaces of \SRO{} projected along $k_z$ under \textbf{(a)} zero stress and \textbf{(b)} $\varepsilon_{zz} = -0.025$. The width of the lines indicates the warping of the Fermi surface along $k_z$. The dashed green line is the 2D zone boundary pf the RuO$_2$ sheet.} 
\label{fig:DFT_FSs}
\end{figure}

\begin{figure}[ptb]
	\centering
	\includegraphics[width=85mm]{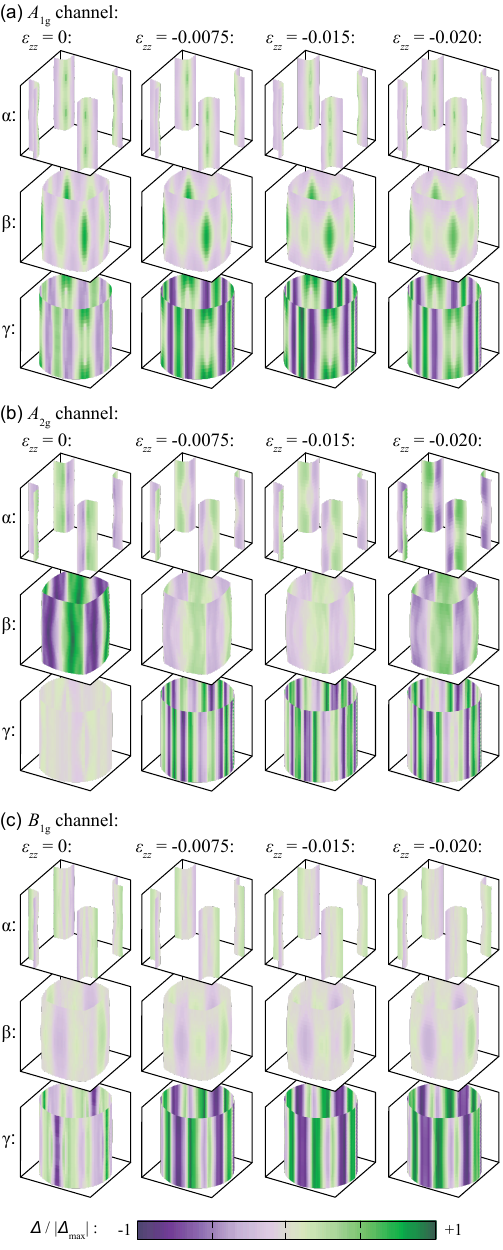}
	\caption{Gap structures at $J/U = 0.15$ for, from top to bottom, the $A_{1g}$, $A_{2g}$, and $B_{1g}$ channels, at the indicated strains. For each channel, the top, middle, and bottom rows show the gap on the $\alpha$, $\beta$, and $\gamma$ sheets, respectively.}
	\label{fig:OPs_compare}
\end{figure}

Uniaxial stress was applied using piezoelectric-driven apparatus \cite{Hicks14_RSI, Barber19_RSI}, and precision in sample mounting is important because \SRO{} is much more sensitive to in-plane than $c$-axis uniaxial stress: $T_\text{c}$ decreases by $0.13$~K under a $c$-axis stress of $\sigma_{zz} = -3.0$~GPa, but increases by $0.13$~K under an in-plane uniaxial stress of only 0.2 GPa~\cite{Steppke17_Science}. 
Applying $c$-axis pressure could generate in-plane stress through bending and/or sample inhomogeneity. 
In a previous experiment~\cite{Kittaka10_PRB}, $c$-axis compression raised $T_\text{c}$ and broadened the transition. 
However, the stress was applied at room temperature, where the elastic limit of \SRO{} is low~\cite{Barber19_RSI}, so these effects may have been a consequence of in-plane strain due to defects introduced by the applied stress.

Samples 2--4 were mounted into two-part sample carriers; that for samples 2 and 3 is diagrammed in Fig.~\ref{fig:setup}(a). 
The purpose was to protect samples from inadvertent application of tensile stress.  
Samples are mounted across a gap between a fixed and a moving portion of part B of the carrier, and can be compressed, but not tensioned, by bringing part A into contact with part B.  
In Fig.~\ref{fig:setup}~(b), we show $T_\text{c}$ of samples $2$ and $3$ versus applied displacement, and the point where parts A and B come into contact and $T_\text{c}$ starts changing is clearly visible. 
For sample 2 the point of contact is rounded on the scale of a few microns, due to roughness and/or misalignment of the contact faces, and in all figures below we exclude data points that we estimate to be affected by this rounding. 

The samples were mounted with Stycast 2850. This epoxy layer constitutes a conformal layer that ensures even application of stress~\cite{Hicks14_RSI}.
Photographs of samples 2--4 are shown in Fig.~\ref{fig:setup}(c--e). 
The carrier for sample 4, which has a different design to those used for samples 2 and 3, is shown in Fig.~\ref{fig:setup}(f). 
Where electrical contacts were made, Du-Pont 6838 silver paste annealed at $450^\circ$ for typically 30 minutes was used.
This is longer than usual, in order to penetrate a thin insulating layer deposited during the ion beam milling.

As noted above, samples 1–--3 were mounted in apparatus that had a sensor only of the displacement applied to the sample, while for sample 4 there was also a force sensor. 
Displacement sensors are less reliable as sensors of the state of the sample, because they also pick up deformation of the epoxy that holds the sample. 
In Fig.~\ref{fig:additionalData}(a) the complete set of measurements of $T_\text{c}$ of sample~3, plotted against applied displacement, are shown.
Data points are colored by the order in which they were collected. 
The data drifted leftward over time: stronger compression was needed to reach the same $T_\text{c}$. 
However, the qualitative form of the curve --- initial decrease in $T_{c}$, then a flattening, and then further decrease --- reproduced over multiple stress cycles, and in panel (b) it is shown that the form of the transition was the same before and after application of the strongest compression.

(We attribute the small apparent shift in $T_\text{c}$ to an artefact of inadvertent mechanical contact between the stress cell and inner vacuum can of the cryostat.) 
We therefore conclude that the sample deformed elastically and that it was the epoxy holding the sample that was compressed non-elastically; plastic deformation has previously been observed to broaden the superconducting transition~\cite{Taniguchi15_JPSJ} of \SRO{}. 
In Fig.~\ref{fig:temp_sweeps}, in the main text, we show only the data taken after the epoxy was maximally compressed.
Force versus displacement data for sample 4 are shown in Fig.~\ref{fig:additionalData}(c–--d), and here it can be seen that there was very substantial non-elastic compression of the epoxy. 
As with sample 3, the shape of the superconducting transition in the \SRO{} was the same before and after application of large stress. 
Over regions where the sample and epoxy deformed elastically, the combined spring constant was 1.45~N/$\mu$m. 
The spring constant of the flexures in the carrier, on the other hand, is calculated to be $\sim$0.03~N/$\mu$m, meaning that almost all of the applied force was transferred to the sample.

\subsection{Electronic structure calculations}
The calculated Fermi surfaces of unstressed \SRO{} and under interlayer compression, including the warping along $k_z$ and the Fermi velocities, are shown in Fig.~\ref{fig:DFT_FSs}. The $\beta$ sheet is the most strongly warped both at zero stress and at $\varepsilon_{zz} = -0.025$.

\subsection{Estimate of $c$-axis strain by resistivity}
In the DFT calculation, the increase in the DOS reaches $20~\%$ at $\epsilon_{zz} \approx -0.017$, around $70~\%$ of the way to the Lifshitz transition. 
The $c$-axis resistivity also allows an estimate of how close we are to the Lifshitz point. 
$c$-axis conductivity is proportional to the square of the amplitude of $k_z$ warping. 
We observe $\rho_{zz}$ to fall by 14--17\% between $\sigma_{zz}=0$ and $-1.0$~GPa, implying an increase in warping amplitude of 8--10\%. 
The $\beta$ sheet is the most strongly warped, and in the DFT calculations, its warping - its area projected on the $k_x-k_y$ plane - is 85\% larger at the Lifshitz transition than at zero strain. 
Therefore, if warping increases linearly with $c$-axis compression the Lifshitz transition is expected to occur between $-8$ and $-11$~GPa.
In practice, the dependence on compression is likely to be super-linear, so we believe that taking into account the various pieces of evidence an estimate that we are approximately half way to the transition point is reasonable.

\subsection{Calculated gap structure in other channels}

In Fig.~\ref{fig:OPs_compare} the calculated gap structures in the $A_{1g}$, $A_{2g}$, and $B_{1g}$ channels are shown. 
[The $B_{2g}$ gap structures are shown in Fig.~\ref{fig:Eigs}(e).] 
$c$-axis compression favors large gaps on the $\gamma$ sheet in all channels. 
In the $A_{2g}$ channel, this shift occurs as a first-order change in gap structure between $\epsilon_{zz} = 0$ and $\epsilon_{zz} = −0.0075$. 
At the largest compression reached, gap weight in the $A_{2g}$ channel shifts back away from the $\gamma$ sheet, as it does in the $B_{2g}$ channel. 
This does not occur in the $A_{1g}$ and $B_{1g}$ channels.

\subsection{Details of the weak-coupling calculation}

The tight-binding Hamiltonian from Eq.~\eqref{eq:Hamiltonian} takes the form
\begin{equation}
\mathcal{H}_s(\boldsymbol{k}) = \begin{pmatrix}
\varepsilon_{AA}(\boldsymbol{k}) & \varepsilon_{AB}(\boldsymbol{k}) - i s\eta_1 & + i\eta_2 \\
\varepsilon_{BA}(\boldsymbol{k}) + i s\eta_1 & \varepsilon_{BB}(\boldsymbol{k}) & - s\eta_2 \\ - i\eta_2 &  -s\eta_2 & \varepsilon_{CC}(\boldsymbol{k})
\end{pmatrix},
\label{eq:HamiltonianMat}
\end{equation}
where we used the Ru orbital shorthand notation $A=xz$, $B=yz$, $C=xy$, and where $\bar{s} = -s$ ($s$ being spin). In Eq.~\eqref{eq:HamiltonianMat} the energies
$\varepsilon_{AB}(\boldsymbol{k})$ account for intra-orbital ($A = B$) and inter-orbital ($A \neq B$) hopping, and $\eta_1$, $\eta_2$ parametrize the spin-orbit
coupling. We define $\varepsilon_{AA}(\boldsymbol{k}) = \varepsilon_{\mathrm{1D}}(k_x, k_y, k_z)$, $\varepsilon_{BB}(\boldsymbol{k}) =
\varepsilon_{\mathrm{1D}}(k_y, k_x, k_z)$, and $\varepsilon_{CC}(\boldsymbol{k}) = \varepsilon_{\mathrm{2D}}(k_x, k_y, k_z)$, and we retain the following terms
in the matrix elements:
\begin{widetext}
\begin{align}
\varepsilon_{\mathrm{1D}}(k_{\parallel}, k_{\perp}, k_z) &= -\mu_{\mathrm{1D}} - 2t_1 \cos(k_{\parallel}) - 2t_2 \cos(k_{\perp}) - 4t_3 \cos(k_{\parallel}) \cos(k_{\perp}) \nonumber \\
&\hspace{10pt} -  8t_4 \cos(k_{\parallel}/2) \cos(k_{\perp}/2) \cos(k_z/2) - 2t_5 \cos(2k_{\parallel}) - 4t_6  \cos(2 k_{\parallel}) \cos(k_{\perp}) - 2t_7 \cos(3k_{\parallel}), \label{eq:1DHopping} \\
\varepsilon_{\mathrm{2D}}(\boldsymbol{k}) &= - \mu_{\text{2D}} - 2\bar{t}_1 \left[ \cos(k_x) + \cos(k_y) \right] - 2\bar{t}_2 \left[ \cos(2 k_x) + \cos(2 k_y) \right] - 4\bar{t}_3 \cos(k_x) \cos(k_y)  \nonumber \\
&\hspace{10pt} -4 \bar{t}_4 \left[ \cos(2 k_x) \cos(k_y) + \cos(2 k_y) \cos(k_x) \right]  - 4\bar{t}_5 \cos(2k_x) \cos(2k_y) \label{eq:2DHopping} \\
&\hspace{10pt} - 4\bar{t}_6 \left[ \cos(3 k_x) \cos(k_y) + \cos(3 k_y) \cos(k_x) \right] - 2\bar{t}_7 \left[ \cos(3 k_x) + \cos(3 k_y) \right] \nonumber \\
&\hspace{10pt} - 8\bar{t}_8 \cos(k_z/2) \cos(k_x/2) \cos(k_y/2), \nonumber \\
\varepsilon_{AB}(\boldsymbol{k}) &= - 8 \tilde{t} \sin(k_x/2)\sin(k_y/2) \cos(k_z/2). \label{eq:offHopping} 
\end{align}
\end{widetext}
Here the first Brillouin zone is defined as $\text{BZ} = [-\pi,\pi]^2\times [-2\pi,2\pi]$. For the four values of $c$-axis compression $\varepsilon_{zz} = 0,~-0.0075,~-0.015,~-0.020$ we extract
the entire set of parameters from DFT calculations consistent with Fig.~\ref{fig:DFT}; see Table~\ref{tab:Parameters}.
\begin{table*}
\centering
\caption{Tight-binding parameters (in units of meV) used for Eqs.~\eqref{eq:HamiltonianMat}, \eqref{eq:1DHopping}, \eqref{eq:2DHopping}, and \eqref{eq:offHopping}, yielding Fig.~\ref{fig:Eigs}(a).}
\resizebox{\textwidth}{!}{%
\begin{tabular}{p{1.1cm} p{0.8cm} p{0.8cm} p{0.8cm} p{0.8cm} p{0.8cm} p{0.8cm} p{0.8cm} p{0.8cm} p{0.8cm} p{0.8cm} p{0.8cm} p{0.8cm} p{0.8cm} p{0.8cm} p{0.8cm} p{0.8cm} p{0.8cm} p{0.8cm} p{0.8cm} p{0.8cm}} 
\hline \hline
$\lvert \varepsilon_{zz} \rvert$ & $\mu_{\text{1D}}$ & $t_1$ & $t_2$ & $t_3$ & $t_4$ & $t_5$ & $t_6$ & $t_7$ & $\mu_{\text{2D}}$ & $\bar{t}_1$ & $\bar{t}_2$ & $\bar{t}_3$ & $\bar{t}_4$ & $\bar{t}_5$ & $\bar{t}_6$ & $\bar{t}_7$ & $\bar{t}_8$ & $\tilde{t}$ & $\eta_1$ & $\eta_2$ \\ \hline
$0.00$ & $316$ & $296$ & $53$ & $-16$ & $17$ & $-57$ & $-15$ & $-12$ & $433$ & $370$ & $-6$ & $123$ & $20$ & $14$ & $3$ & $3$ & $-2$ & $9$ & $-51$ & $-51$ \\
$0.0075$ & $294$ & $284$ & $54$ & $-16$ & $18$ & $-56$ & $-15$ & $-10$ & $437$ & $369$ & $-5$ & $122$ & $20$ & $14$ & $3$ & $3$ & $-3$ & $10$ & $-51$ & $-51$ \\
$0.015$ & $273$ & $271$ & $55$ & $-17$ & $19$ & $-55$ & $-15$ & $-9$ & $441$ & $368$ & $-5$ & $121$ & $20$ & $13$ & $3$ & $3$ & $-3$ & $10$ & $-51$ & $-51$ \\
$0.020$ & $259$ & $264$ & $56$ & $-18$ & $20$ & $-55$ & $-15$ & $-8$ & $443$ & $367$ & $-4$ & $120$ & $20$ & $13$ & $3$ & $3$ & $-3$ & $11$ & $-52$ & $-51$ \\ \hline\hline
\label{tab:Parameters}
\end{tabular} }
\end{table*}

For the interactions we use the (on-site) Hubbard--Kanamori Hamiltonian
\begin{widetext}
\begin{equation}
H_I = \frac{U}{2} \sum_{i,a,s\neq s'} n_{i a s} n_{i a s'} +  \frac{U'}{2} \sum_{i, a\neq b,s, s'} n_{i a s} n_{i b s'} + \frac{J}{2} \sum_{i,a\neq b,s, s'} c_{i a s}^{\dagger}
c_{i b s'}^{\dagger}c^{\phantom{\dagger}}_{i a s'}c^{\phantom{\dagger}}_{i b s} + \frac{J'}{2} \sum_{i,a\neq b,s\neq s'} c_{i a s}^{\dagger}
c_{i a s'}^{\dagger}c^{\phantom{\dagger}}_{i b s'}c^{\phantom{\dagger}}_{i b s}, 
\label{eq:Interaction}
\end{equation}
\end{widetext}
where $i$ is site, $a$ is orbital, and $n_{i a s} = c_{i a s}^{\dagger}c_{i a s}$ is the density operator. We further assume that $U' = U-2J$ and $J' = J$~\cite{DagottoEA11}. In the weak-coupling limit this leaves $J/U$ as a single parameter fully characterizing the interactions.

\begin{figure}[ptb]
	\centering
	\subfigure[]{\includegraphics[width=0.45\linewidth]{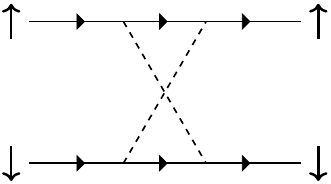}} \quad 
	\subfigure[]{\includegraphics[width=0.45\linewidth]{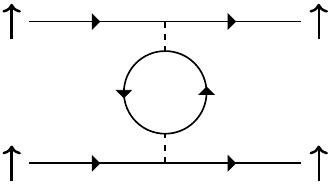}} 
	\caption{Second-order diagrams taken into account in $\Gamma$ in \textbf{(a)} the even-parity channel, and \textbf{(b)} the odd-parity channel. The vertical arrows denote pseudo-spin, and the dashed lines contain all the terms of Eq.~\eqref{eq:Interaction}. The approach is asymptomatically exact in the weak-coupling limit, $U/t \to 0$.}
	\label{fig:Diags}
\end{figure}

In the linearized gap equation~\eqref{eq:GapEquation} the (dimensionless) two-particle interaction vertex $\bar{\Gamma}$ is defined as~\cite{RaghuEA10}
\begin{equation}
\bar{\Gamma}(\boldsymbol{k}_{\mu}, \boldsymbol{k}_{\nu}) = \sqrt{\frac{\rho_{\mu} \bar{v}_{\mu}}{v_{\mu}(\boldsymbol{k}_{\mu}) } } \Gamma(\boldsymbol{k}_{\mu}, \boldsymbol{k}_{\nu}) \sqrt{\frac{\rho_{\nu} \bar{v}_{\nu}}{v_{\nu}(\boldsymbol{k}_{\nu}) } \vphantom{\frac{\rho_{\mu} \bar{v}_{\mu}}{v_{\mu}(\boldsymbol{k}_{\mu}) }}   },
\label{eq:gmatrix}
\end{equation}
where $\rho_{\mu} = \lvert S_{\mu} \rvert /[\bar{v}_{\mu} (2\pi)^3 ]$ is the density of states, and $1/\bar{v}_{\mu} = \int_{S_{\mu}} \mathrm{d} \boldsymbol{k} / \left(\lvert S_{\mu} \rvert
v_{\mu}(\boldsymbol{k}) \right)$. Here, $\Gamma$ is the irreducible two-particle interaction vertex which to leading order retains the diagrams shown in Fig.~\ref{fig:Diags}.

An eigenfunction $\varphi$ of Eq.~\eqref{eq:GapEquation} corresponding to a negative eigenvalue $\lambda$ yields the superconducting order parameter
\begin{equation}
\Delta(\boldsymbol{k}_{\mu}) \sim  \sqrt{\frac{v_{\mu}(\boldsymbol{k}_{\mu})}{\bar{v}_{\mu}\rho_\mu}  } \varphi(\boldsymbol{k}_{\mu}).
\label{eq:GapFunction}
\end{equation}
In the chosen pseudo-spin basis each eigenvector $\varphi$ belongs to one of the ten irreducible representations of the crystal point group $D_{4h}$~\cite{SigristUeda91, RaghuEA10}.

\bibliography{draft_220102_arxiv.bib}

\end{document}